# Pulsatility index as a diagnostic parameter of reciprocating wall shear stress parameters in physiological pulsating waveforms

**Idit Avrahami[1], Dikla Kersh[2] and Alex Liberzon[2]**

[1] Department of Mechanical Engineering and Mechatronics, Ariel University, Israel

[2] School of Mechanical Engineering, Tel Aviv University, Tel Aviv 69978, Israel

**Corresponding author:**

Prof. Alex Liberzon, PhD

School of Mechanical Engineering

Tel Aviv University

Tel Aviv 69978, ISRAEL

E-mail: alexlib@eng.tau.ac.il

# Pulsatility index as a diagnostic parameter of reciprocating wall shear stress parameters in physiological pulsating waveforms

Idit Avrahami[1], Dikla Kersh[2] and Alex Liberzon[2]

**Abstract** Arterial Wall shear stress (WSS) parameters are widely used for prediction of the initiation and development of atherosclerosis and arterial pathologies. Traditional clinical evaluation of arterial condition relies on correlations of WSS parameters with average flow rate (Q) and heart rate (HR) measurements. We show that for pulsating flow waveforms in a straight tube with flow reversals that lead to significant reciprocating WSS, the measurements of HR and Q are not sufficient for prediction of WSS parameters. Therefore, we suggest adding a third quantity - known as the pulsatility index (PI) - which is defined as the peak-to-peak flow rate amplitude normalized by Q. We examine several pulsating flow waveforms with and without flow reversals using a simulation of a Womersley model in a straight rigid tube and validate the simulations through experimental study using particle image velocimetry (PIV). The results indicate that clinically relevant WSS parameters such as the percentage of negative WSS (P[%]), oscillating shear index (OSI) and the ratio of minimum to maximum shear stress rates (*min/max*), are better predicted when the PI is used in conjunction with HR and Q. Therefore, we propose to use PI as an additional and essential diagnostic quantity for improved predictability of the reciprocating WSS.
**Keywords** wall shear stress, pulsatility index, oscillating shear index, hemodynamics

# 1 Introduction

Arterial wall shear stresses (WSS) are known as a main regulator of endothelial cell function and vascular structure and condition. Numerous *in-vitro* and *in-vivo* studies have shown that vascular regions with disturbed flow accompanied by turbulent flow, low, oscillatory or instantaneous negative WSS and high WSS gradients are strongly correlated with vascular pathologies, cardiovascular diseases and disorders [1-3].

Vascular remodeling and dysfunction were correlated with reciprocating WSS parameters such as the oscillatory shear index (OSI) [4], WSS gradients, percentage of negative wall shear stress (P[%]) or the ratio of minimum to maximum shear stress rates (*min/max*) [5-7]. WSS parameters were shown to have positive correlation with a large variety of vascular dysfunction mechanisms such as reduction in Nitric Oxide (NO) production, endothelial cell activation and platelet adhesion leading to vascular pathologies, including intimal hyperplasia [8], atherosclerosis [9], early plaque deposition and arterial thrombogenesis [10-12], plaque progression [4, 6] and vulnerability [13], arterial aneurysm location [14, 15], growth [16] and rupture [17, 18],

WSS parameters were also found as indicators for advanced congestive heart failure, where the combination of reduced cardiac output compensated by increased HR promotes retrograde flow and negative WSS in the aorta over a major part of the cardiac cycle [20].

Clinical evaluation of time-dependent WSS *in-vivo* is not yet the part of routine cardiovascular exams. This is mainly because it requires accurate measurements of either the direct flow velocity gradient near the wall - using complicated MRI or Doppler ultrasound technique [21-23], or the indirect estimates using intrusive pressure gradient measurements. Pressure gradient measurements in conjunction with the classical Womersley solution for pulsatile flow in rigid tubes [24] may provide time-dependent WSS according to:

$$\text{WSS} = \tau(t) = \sum_{n=0}^{N} -k_{sn} \frac{D}{2(i^{3/2}\alpha_n)} \cdot \left( \frac{J_1(i^{3/2}\alpha_n)}{J_0(i^{3/2}\alpha_n)} \right) e^{i\omega nt} \qquad (1)$$

where $j$ is the imaginary unit, D is the tube inner diameter; $J_0$ and $J_1$ are the zero- and first-order Bessel functions of the first kind. The solution is a sum of *N+1* harmonics of the basic frequency $\omega_0$, where $k_{sn}$ are the amplitudes of oscillatory pressure gradient and $\alpha_n = 0.5D(\omega n/v)^{1/2}$ is the Womersley number of the $n^{th}$ harmonic ($\omega_n = \omega n$), respectively. $v$ is the kinematic viscosity.

A third option is to estimate *in-vivo* time-dependent WSS using the Dopller ultrasound flow waveform measurements and "Inverse Womersley method" [25]. This method allows to extracts the time-dependent WSS from the measured waveform Q(t), by estimating Womersley velocity profiles (*u(t,r)*) based on the time-dependent flow in a straight tube, and extracting the shear stress from the estimated velocity gradients:

$$\text{WSS} = \tau_{wall} = \mu \frac{du}{dr}\Big|_{r=R} \qquad (2)$$

where $\mu$ is dynamic viscosity.

Thus, it should be possible, at least in theory, to estimate the time dependent WSS by applying either straight Womersley method on the time-dependent measured pressure or the inverse Womersley method on the time-dependent measured flow waveform. However, these methods are still too complex to be applied in clinical practice since they require accurate time-resolved measurements. In addition, the accuracy of these methods relies on the validity of Womersley solution, approximating the blood vessel to a straight uniform and rigid tube. The reported discrepancies of WSS measured using the inverse Womersley method as compared to more complex non-linear approximations, are in the range of 23% [26].

Given the above, it is desirable to establish clinical evaluation based on simpler and more accessible measurements that would provide improved estimation of the reciprocating WSS parameters, such as OSI, P[%] or *min/max*.

Some studies tried to correlate reciprocating WSS parameters with the basic integral parameters of averaged or maximal flow rate (*Q*) and heart rate (HR), which are the simplest to measure. For example, Finol et al. [14] found that the maximal values of non-dimensional mean WSS and non-dimensional WSS gradient increase with the Reynolds number (non-dimensional representation of the flow rate according to $\mathrm{Re} = 4Q/\pi D\nu$). Other researchers [27, 28] have shown correlations between increased WSS magnitude and increased Womersley number (non-dimensional representation of the HR according to $\alpha = 0.5D(\omega/\nu)^{1/2}$) during exercise . Increased HR is considered a major risk factor for cardiovascular disease [28, 29], and when medically treated, was found to reduce heart failure [30].

However, as we show in this study, the correlations based only on HR and Q cannot be sufficient due to the nature of the pulsating flow waveforms. In some flow cases for the same HR (Womersley number) and averaged flow rate (Reynolds number) the flow waveform and associated local WSS parameters, may differ drastically due to reverse flow rate during a portion of the cardiac cycle. For example, the aortic wave steepening downstream of the aorta, caused by wave reflection and diameter reduction results in different flow waveforms at different locations along the artery, therefore different WSS parameters for the same HR and Q [31, 32]. Under such conditions, the peak of the maximum flow rate increases but it is compensated over the cardiac cycle by the stronger reverse (negative) flow rate, causing a lower net positive flow rate.

It is important to distinguish between the reverse flow which is *local* (with negative velocity values only in the near wall region but with the overall positive flow rate), and *global reverse flow*, when the flow rate at the given cross-section of the artery is negative ($Q < 0$). In both local and global reverse flow cases there is negative WSS, but on very different time scales [33].

We suggest that the pulsatility index (PI) could be used as a third dimensionless quantity to predict WSS parameters. PI was proposed by Goslin and King [34] to measure the variability of blood velocity in a vessel, and it is defined as the difference between the peak systolic flow and minimum diastolic flow rates divided by the mean flow rate $\left( \text{PI} = (Q_{max} - Q_{min})/Q_{mean} \right)$. PI is readily accessible *in-vivo* [35]. It has been shown to correlate with presence of arterial stenosis [36], and been recently recognized as an important factor in aneurysmal flows [7, 15, 18, 37] or acute intracerebral hemorrhage [38]. In a different context, the PI is called reverse/forward flow index (measured in percent, see Hashimoto and Ito [39], for instance) and it shows correlations with the pulse pressure amplification and arterial stiffness.

In this study we present in a simulation and verify experimentally that in a straight tube, addition of PI allows better prediction of WSS parameters than Q and HR alone and would potentially improve prediction of negative WSS during the cardiac cycle. To the best of our knowledge, prediction of WSS parameters utilizing PI, as developed in this study was not published elsewhere.

The manuscript is organized as follows. Section 2 presents the methods used in the study, namely the simulation based on the Womersley model and an experimental study using particle image velocimetry (PIV). Section 3 presents the results of the observed correlations and followed by discussion of the results along with the limitations of the present studies and the future perspectives of this research.

## 2   Materials and Methods

In order to demonstrate the capabilities of PI in improving prediction of WSS parameters for waveforms with global reverse flow and negative WSS and distinguish those from cases of negative WSS with positive flow rate, we demonstrate in this manuscript the correlation of WSS parameters with Reynolds, Womersley and PI for 14 representative pulsating waveforms (run cases). The waveforms were chosen such that they could be realized in our experimental setup and implemented in both simulations and the experiment. Chronologically, the experiments were performed before the simulation. The experimentally accessible waveforms were analyzed to obtain the amplitudes and phases of the flow rate waveforms and simulated using inverse Womersley solution (32). WSS was estimated using Eq. (2).

Based on time-dependent WSS, $\tau(t)$, we quantify the WSS-related parameters (using Eq.3), shown to correlate with clinical evidence of vascular dysfunction, namely a) the ratio of min/max WSS, b) the time interval of the negative WSS during the cycle (P[%]), and c) the measure of overall oscillation (the so-called oscillation shear index, OSI). The first two parameters, min/max and P[%] were defined by Gharib and Beizaie [11] and were shown to be correlated with the clinical findings. OSI is an index defined by Ku et al. [7] to describe the degree of deviation of the WSS from its average direction. These parameters are defined as follows:

$$\min/\max = \left|\frac{\tau_{min}}{\tau_{max}}\right|$$

$$P[\%] = \frac{t|_{\tau<0}}{T} \qquad (3)$$

$$OSI = \frac{1}{2}\left(1 - \int_0^T \tau(t)\,dt \bigg/ \int_0^T |\tau(t)|\,dt\right)$$

where $\tau_{min}$ and $\tau_{max}$ are the minimal and maximal values of time-dependent WSS in the cycle, $t|_{\tau<0}$ is the total time during which WSS is negative, and $T$ is the time period of the pulsation (inverse of the HR).

A custom-design flow test rig was developed to create controllable pulsatile flows, as shown in Figure 3a. Pulsatile flow waveforms were developed by three DC voltage-driven computer-controlled gear pumps: two pumps connected in parallel drove the flow in the streamwise (forward) direction, and the third pump was installed in the inverse direction to control the flow reversals (Figure 3a). The flow was created in a distensible tube (made of Tygon B-44-4X, Saint Gobain, 80 cm long, inner diameter of 1.9 cm and wall thickness of 0.32 cm, $L/d \geq 40$, elasticity modulus of 12 MPa, estimated distensibility $1.36 \times 10^{-6}$ Pa$^{-1}$). The experimental cases examined had typical physiological characteristics composed of 30-50% "systolic" (forward) part followed by a "diastolic" component. The cases differ by the mean flow rate, presence/absence of the negative WSS and presence/absence of the total flow reversal (i.e. negative flow rate) during a certain part of the cycle. Examined flow waveforms were with relevant parameters at the relevant range for large vessels with Reynolds number ranged between $Re_{max}$= 70 ÷ 750 (associated with maximal peak systolic flow rate), Womersley number $\alpha$= 6 ÷ 13 and $PI$ = 1.5 ÷ 9 [40]. The dimensionless parameters of each experimental run are listed in Table 1.

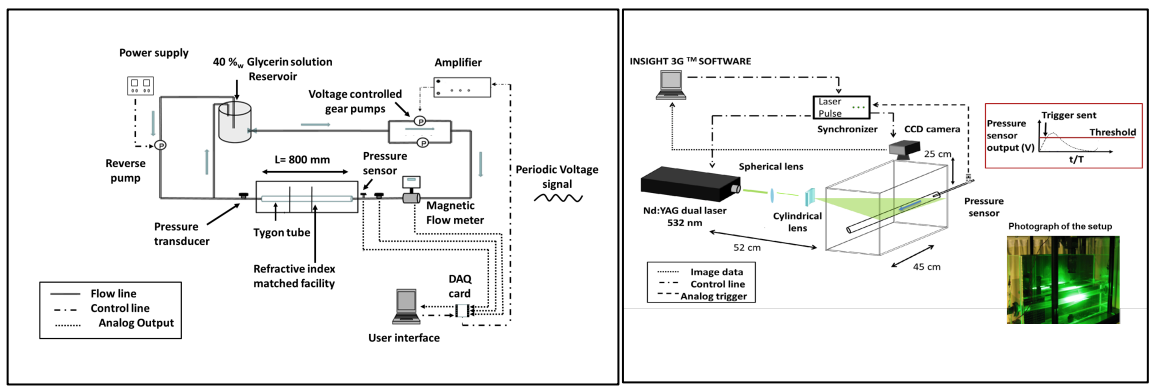

Figure 3: Schematic description of the experimental setup: (a) hydraulic setup, including the computer-controlled two-directional pumping system, elastic tube embedded in a refractive index matched solution, and the pressure/discharge metering system. (b) Schematic view of the phase-averaging PIV measurement system, computer controlled waveform pressure signal used to trigger the PIV acquisition at each cycle.

The measurements included pressure transducers at the inlet and outlet of the tube (EW-68075-02, Cole Parmer), flow rate (magnetic flow meter MAG 1100, Danfoss) and velocity distribution using particle image velocimetry (PIV) as shown in Figure 3b. In order to reduce optical reflections and distortion, the tube was placed in a 800 × 300 × 200 mm glass tank filled with a 60%$_w$ glycerin-water solution, which refractive index matches to the Tygon tube. For the working fluid inside the tube, a 40%*w* glycerin-water was used as a model of blood. Although this fluid is Newtonian and homogenous, it is a commonly used solution to match the blood viscosity [41]. Its dynamic viscosity is 3.72 cP and density is 1.09 g/cm$^3$ as compared to blood with 3-4 cP and 1.06 g/cm$^3$. The small refractive mismatch between the tube and the working fluid resulted with some light reflection from the pipe walls that affect PIV measurement resolution at the near wall region.

We implemented particle image velocimetry (PIV) technique using a dual Nd:YAG laser (532 nm, 120 mJ/pulse, Solo 120XT, New Wave Research), a high-resolution CCD camera (12 bit, 4008×2672 pixels, TSI Inc.). The field of view was located 0.6 m from the inlet, downstream of the inlet region of the tube, in order to assure the fully developed flow at the measurement site. The field of view was set to 60 x 40 mm with a magnification ratio of 15 µm/pixel. The PIV analysis was performed on rectangular windows of 64 x 16 pixels. Silver-coated hollow glass spheres (14 *µm*, 1.05 g/cm$^3$, TSI Inc.) were used as seeding particles. An analog pressure measurement located at the inlet was used to trigger the PIV synchronizer at fixed phase instants. Seven double-laser pulses and PIV images were taken at fixed time instants during each cycle (at t/T= 0, 0.125, 0.25, 0.375, 0.5, 0.625 and 0.75 equally spaced sections of the normalized period). For each of the seven phases during the cycle, 100 images were acquired. Velocity fields were estimated using a commercial software (Insight3G, TSI Inc.) and verified with an open source software (www.openpiv.net) [42]. First, the 50 flow realizations at each phase were averaged to obtain the phase averaged PIV flow realizations. Second, the velocity was averaged along the streamwise (*x*) direction, arriving at the phase averaged velocity profiles, *u(r,t)*. The phase averaged velocity profiles are shown in Figure 4 as symbols and colors are used to distinguish between different phases. In some cases, the velocity profiles overlap at different phases, and only a part of those are shown for the sake of clarity. The flow rate waveforms for each of these cases are shown in Figure 5 (dashed lines).

The experimental results are analysed for quantification of the WSS and its relation to the presence of local (near the wall) or global reverse flow (negative flow rate) during a part of each waveform cycle. The most accurate method is the extended version of inverse Womersley solution. The PIV system provides a set of seven time phases along the cycle but with high spatial resolution in the tube cross-section. We apply constrained non-linear least-squares method to find the best fit of Womersley profiles to the phase-averaged measured velocity profiles, according to:

$$u(r,t) = \text{Real}\left\{ \sum_{n=0}^{N} \frac{4Q_n}{\pi D^2} \left( \frac{j^{3/2}\alpha_n J_0(j^{3/2}\alpha_n) - j^{3/2}\alpha_n J_0\left(j^{3/2}\alpha_n \frac{2r}{D}\right)}{j^{3/2}\alpha_n J_0(j^{3/2}\alpha_n) - 2J_1(j^{3/2}\alpha_n)} \right) e^{j\omega nt} \right\} \quad (4)$$

The non-linear least squares fit provides the amplitudes and phases of flow rate harmonics. Using the amplitudes and phases we evaluate velocity gradients according to Eq. 4 and than calculated WSS according to Eq.(2) and the respective parameters defined in Eq. (3). We then study their correlations with PI, Re and $\alpha$.

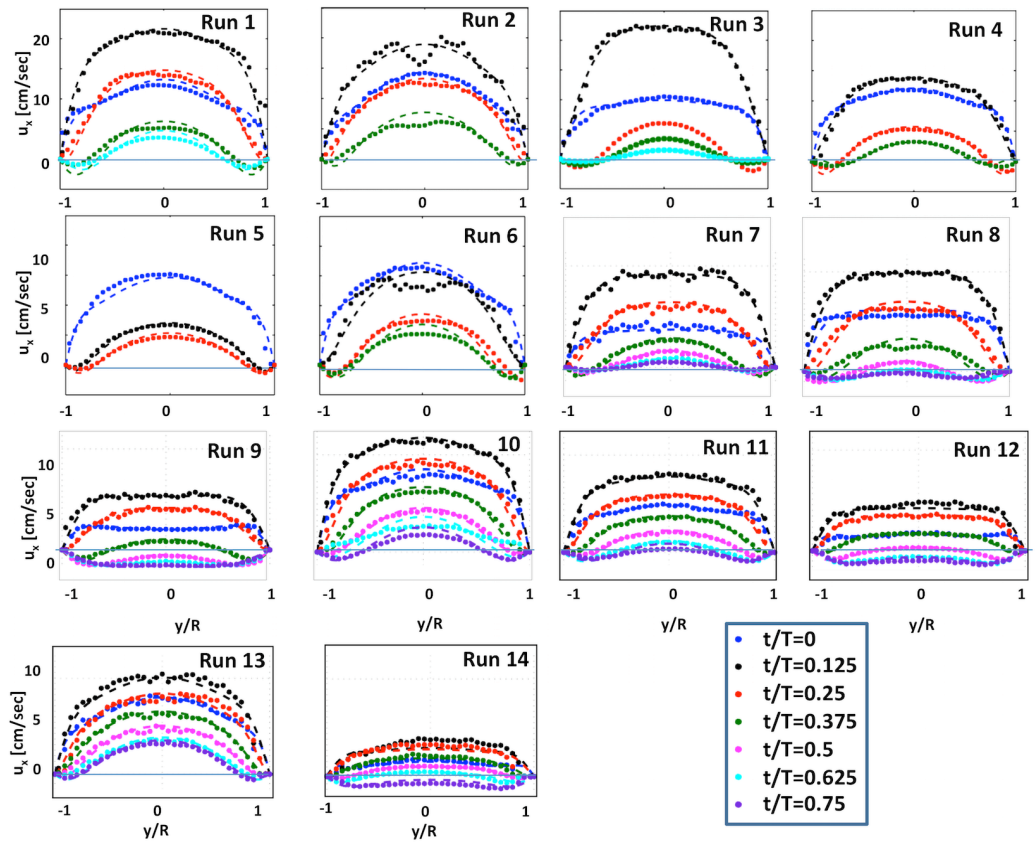

Figure 4: PIV measured velocity profiles (symbols) and the corresponding inverse Womersley solution profiles (dashed-lines) for 14 runs at different time phases (color online).

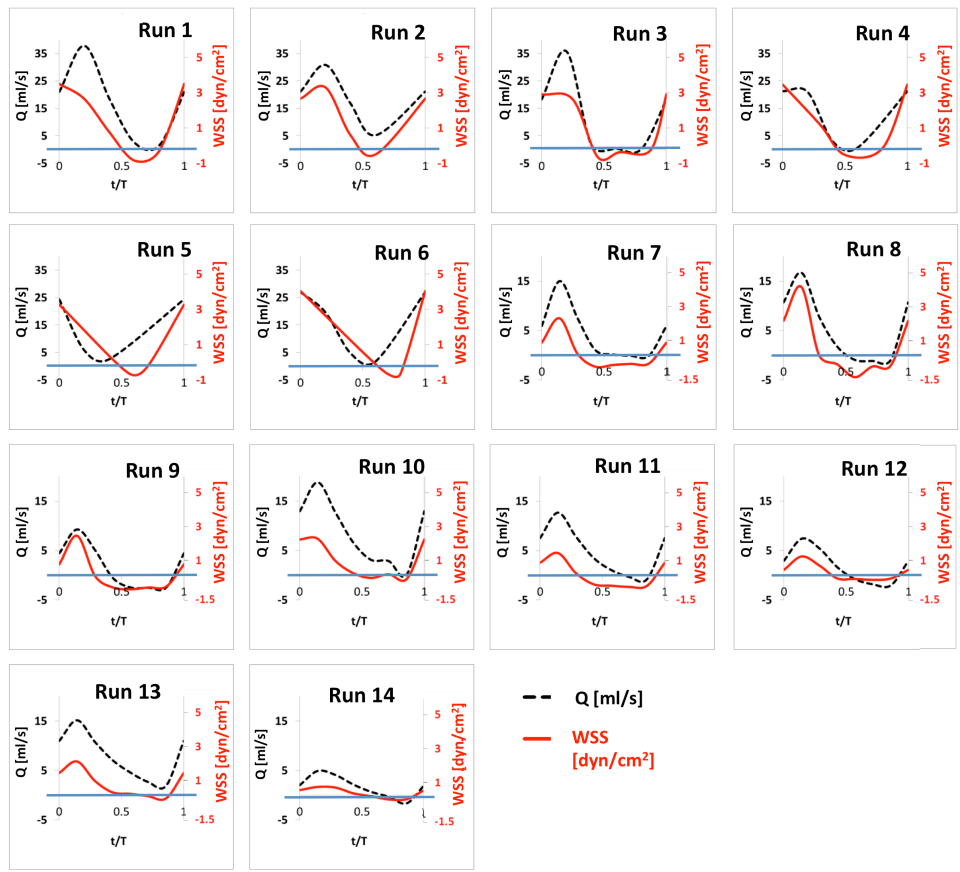

Figure 5: Flow rate (dashed black lines, left axis) and the corresponding WSS values (red solid lines, right axis) as simulated for the 14 cases

Figure 6a shows an example of the analysis for two example cases: one case with only positive flow throughout the entire cycle (Run 10, left) and the second case with a total flow reversal (Run 12, right). It shows the comparison between the flow rate Q waveform obtained directly from integration of PIV velocity profiles in radial direction (symbols with error bars at seven phases) and the time-dependent flow rate obtained from the inverse Womersley method (solid line). Similarly, the WSS estimated derivative of the velocity profiles near the wall as measured by PIV (symbols in Figure 6b) are compared with the WSS obtained from the inverse Womersley method according to Eqs. (2) and (4) (solid lines).

Avrahami et al.

The differences emphasize intrinsic limitations of both the direct WSS measurements from PIV data near the wall and the fact we fit a linear Womersley solution in a rigid tube to the real flow case in a slightly distensible tube. Although it is theoretically possible to include nonlinearities arising from tube's distensibility, curvature, tube-end conditions and nonuniformity using the modified Womersley solution for thin wall elastic tubes [43], this method would significantly complicate the simulations.

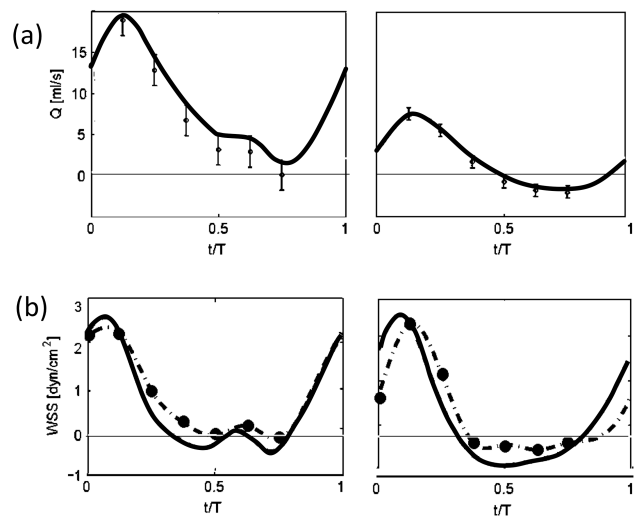

Figure 6: An example of flow waveforms and respective WSS estimate for two representative experimental runs: run #10 with only positive flow throughout the entire cycle (left), and run #12, with total flow reversal (right) (a) Flow rate as a function of normalized time (Q(t/T)). Values integrated from the PIV measurements are shown by symbols and the associated error bars, while curves represent the values evaluated using the inverse Womersley model. (b) Comparison between WSS calculated from Womersley approximation (solid line) and WSS calculated from measured velocity profiles (dashed line with symbols).

## 3  Results

Using the simulation and experimental results, we estimate WSS parameters defined in Eq. 3 and present them as functions of the Reynolds, Womersley, and pulsating index numbers for each waveform, as listed in Table 1.

The main result of this study is shown in Figures 5-7 and summarized in Table 2. In each figure, panels (a-c) present the WSS parameters of interest, namely OSI, P% and min/max, versus each of the three hemodynamic parameters of Reynolds, Womersley and pulsatility index, separately. We add to each figure a best-fit curve, the R-square value of which is listed in Table 2. We limit our fit to the 2$^{nd}$ order polymonial, for the sake of simplicity and robustness. Higher order polynomials, as well as other types of functions have not shown better fit, though increased complexity and noise amplification during the following, predictive step. As can be seen in Figure 7-7(a-c), the WSS parameters do not have clear relations with the separate dimensionless quantities. The exceptions are the decrease of P%, OSI with the increasing Reynolds number and increase of those with the PI.

Next step after analysing the separate dimensionless quantities, is to find a non-linear fit of two dimensionless quantities together and construct the WSS-related parameters as functions of two arguments, for instance of Reynolds and Womersley numbers (blue squares in panels d). These attempts are compared in Figures 5-7 (d) to the models of Re and PI as the arguments (magenta triangles) and also the three-variable model that we focus on in this work (red circles). The models are constructed by non-linear least squares fit of the 2$^{nd}$ order polynomials in the form of $\hat{Y} = c_0 + c_1 \text{Re} + c_2 \text{Re}^2 + c_3 \alpha + c_4 \alpha^2 + c_5 PI + c_6 PI^2$ , where $c_i$ are the coefficients obtained from a least square fit and $\hat{Y}$ represents the prediction of one of the WSS parameters. Figures 5-7 (d) present the regression between the measured WSS parameter and the estimated values based on the obtained coefficients. The thin straight line in panels (d) emphasizes a theoretically perfect prediction line at 45 degrees. The scatter distribution of the predicted parameter values in the graphs and the R-square values (least square error normalized by the variance of the parameter) listed in Table 2 strengthen the central result of this work. Inclusion of the pulsatility index in the list of significant hemodynamic parameters (in addition to the Reynolds and Womersley numbers) that characterize the Womersley-like flow in straight and rigid pipes provides the best prediction for all three WSS-parameters.

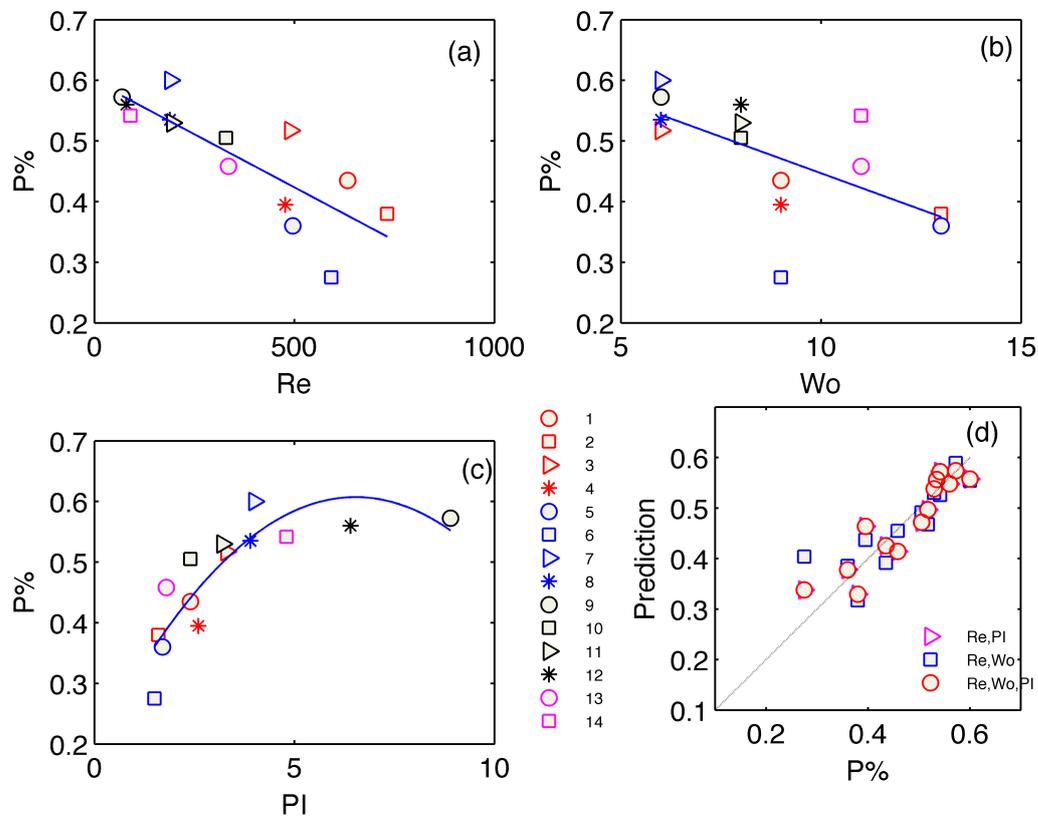

**Figure 1:** **Scatter plot of *P%* versus Reynolds (a), Womersley (b) and PI (c) parameters. Regression between the measured OSI values and predicted *P%* values (d), based on the non-linear least-square fit of a second order multi-variable polynomial function f(Re,Wo,PI) . The cases in the graphs are marked according to their run number (as listed in Table 1) (see legends at the bottom).**

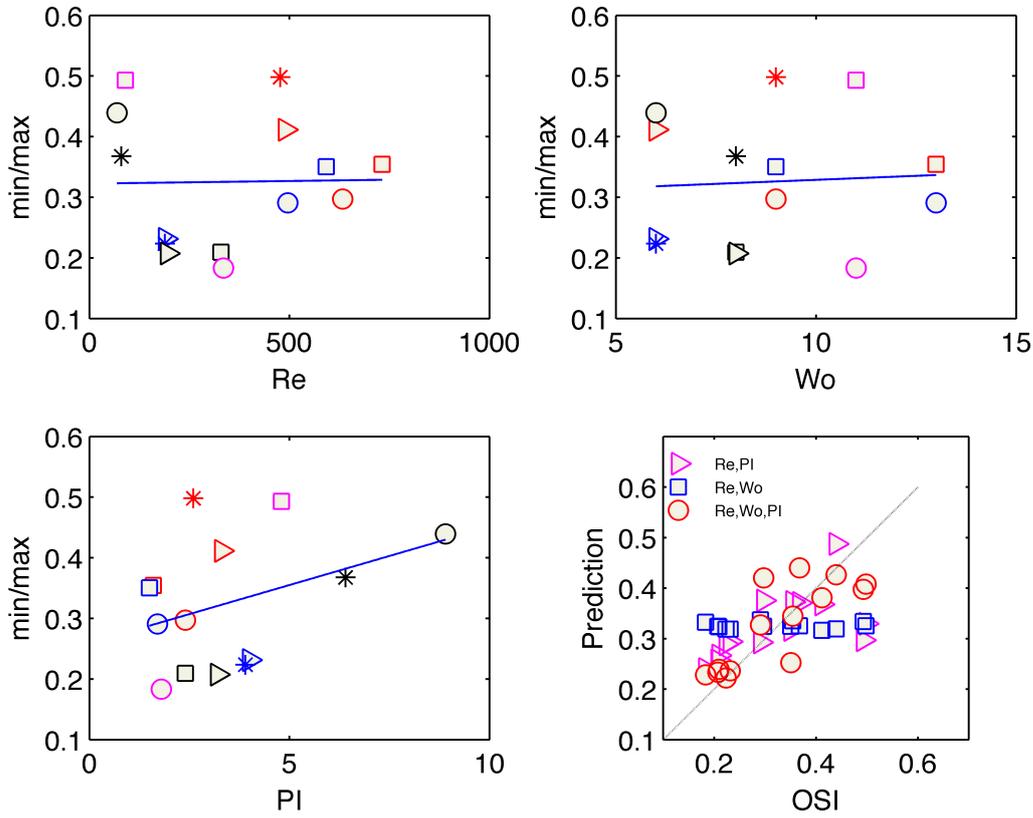

**Figure 2:** **Scatter plot of *min/max* versus Reynolds (a), Womersley (b) and PI (c) parameters. Regression between the measured OSI values and predicted *min/max* values (d), based on the non-linear least-square fit of a second order multi-variable polynomial function f(Re,Wo,PI) .**

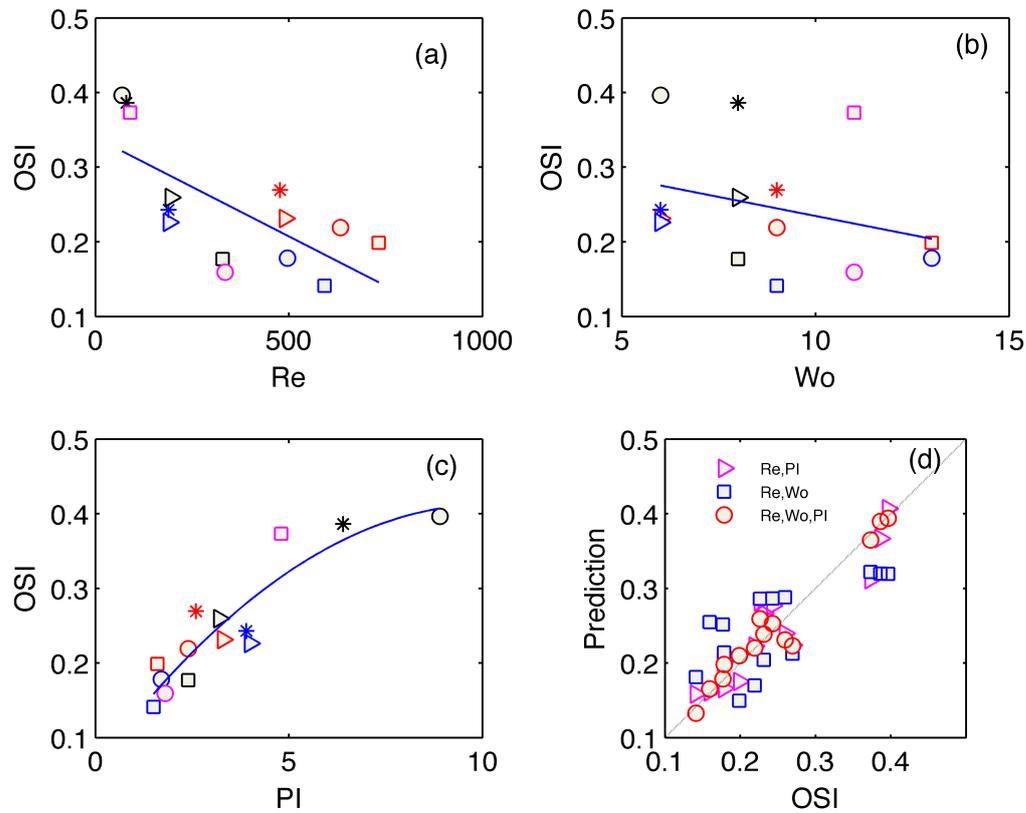

Figure 7: Scatter plot of *OSI* versus Reynolds (a), Womersley (b) and PI (c) parameters. Regression between the measured OSI values and predicted P% values (d), based on the non-linear least-square fit of a second order multi-variable polynomial function f(Re,Wo,PI). The cases in the graphs are marked according to their run number (as listed in Table 1) (see legends at the bottom).

## 4   Summary and Discussion

In this study we show that the use of the flow rate (Q) and heart rate (HR) is not sufficient for prediction of WSS parameters, and we suggest to improve the diagnostic evaluation by adding a third dimensionless parameter, the pulsatility index (PI).

Figure 7-7 and Table 2 demonstrate the limitations of non-dimensional Reynolds and Womersley numbers alone and the benefit of adding the PI for better prediction of WSS parameters.

The results show that Re and $\alpha$ separately or together have poor correlation with the WSS parameters, with the Womersley number lacking correlation as a single hemodynamic characteristic to the WSS parameters. This is expected, as the definitions of the WSS parameters are mostly normalized to the total cycle period and, therefore, the basic frequency has very little effect on the WSS waveform. An exception is the percentage of the negative wall shear stress, which slightly decreases with the increasing Womersley number.

When coming in pairs, the correlation of the dimensionless parameters does not improve dramatically in predicting WSS parameters. Only when the three parameters are used together, the correlations are significantly improved.

Among the three parameters of interest, *min/max* seems to be the parameter, which is poorly predicted by the hemodynamic quantities. Apparently this is due to the nature of its definition of the ratio of two peak values unrelated to the underlying fluid dynamics. High (or low) peaks can appear in weak (or strong) flows with weak or strong pulsations or with high (low) flow rates. Thus, prediction of *min/max* is problematic and if this parameter is of interest, direct measurement of the peaks is required.

The *P[%]* and most importantly, the OSI parameters are very well related to the $2^{nd}$ order polynomial multi-variable model that involves the three hemodynamic parameters. Large OSI and *P[%]* values for large PI values imply that the global flow reversal is a major contributing mechanism to WSS variability during the cycle as well as to the duration of the negative WSS. The results undoubtedly reflect the need for an additional, third, parameter that influences WSS parameters in pulsating flows independently of the HR or Q.

The application seems to be straightforward: by means of sampling flow waveform (e.g. using ultrasound Doppler) at several phases during the cardiac cycle, one can readily obtain PI in addition to the traditional HR and flow rate measurements. Joint analysis of these three values will then provide the necessary prediction of negative WSS and severity of associated WSS-related parameters.

Addition of PI to the measurement list may improve significantly prognosis and prediction of vascular disease development due to vascular remodeling and variations in flow pulsatility. Future research could implement clinically the above suggestion and investigate correlations of local PI, HR and Q, extracted from Doppler measurements with vascular dysfunctions.

### 4.1 Study limitations and further research

The use of the inverse Womersley method in the present study considers a simplified geometry of a straight long circular tube, and thus does not take into account complicated anatomical geometries, arterial stenoses, arterial curvatures, bifurcations and nonuniform cross sections – which may contribute vortical or helical 3D flow, skewed flow profiles, separation regions, etc. in addition, wall motion may affect WSS parameters in up to 15%. Therefore, the described analytical method approximates flow profiles and corresponding WSS only for the specific case described. Such simplifications were used in other studies (e.g. [26, 44]).

Another limitation relates to the working fluid. The working fluid used as a model for blood is not perfectly matched with the Tygon tube. The small refractive mismatch resulted with some light reflection from the pipe walls. Static and dynamic calibrations were implemented to ensure accurate measurement. However, measurement resolution at the near wall region was limited.

Due to limited time resolution of the PIV system, the velocity profiles were measured only at seven time instants during the cycle. It could be beneficial to achieve a better accuracy with the time-resolved PIV systems [34]. Nevertheless, using the high *spatial* resolution of the PIV method and the inverse Womersley method, we overcame this limitation and succeeded identifying an important improvement in correlation between the WSS and WSS-related parameters, and nondimensional numbers, using the addition of the pulsatility index.

The inverse Womersley method is based on Womersley's simplified model of a straight, rigid and infinitely long tube. It was shown in previous studies that the goodness of fit of the models may be affected by the geometry and boundary conditions [45], Sahtout and Ben Salah [46] have shown that Womersley linear theory is accurate enough for the long rigid vessel with no taper or distensibility. Yet, the present experimental case is neither rigid, nor a straight infinite pipe with a fixed inner diameter. The Tygon tube, clamped at its end points was slightly curved (due to its distensibility) ts internal diameter is known only approximately along the length Moreover, the pipe is elastic and reacts to the pressure pulses by small, yet non-negligible changes in diameter. These local changes are different for different pressure pulses and flow rates. Nonlinearities arising from distensibility and/or tubes nonuniformity may affect the accuracy of the results by up to 18% [26].

Although the model we used for rigid straight tubes has limited accuracy, it is the simplest model that allows for extending our analysis to the full cardiac cycle, as well as to explain in future work cases that we did not examine in this study.

Recent studies show the feasibility of using the inverse Womersley method in conjunction with the Doppler ultrasound measurements to assess the flow rates and the velocity profiles in rigid tubes [44]. Our contribution is the comprehension of the inconsistency in the analysis based on the mean flow rate and the frequency (Womersley number) only, and the necessity to include the third parameter, the pulsatility index.

## 5    Acknowledgements

The authors are thankful to Roman Povolotsky for the technical support and design of the experimental apparatus. This study was supported in part by a grant from the Nicholas and Elizabeth Slezak Super Center for Cardiac Research and Biomedical Engineering at Tel Aviv University and partially by the Israel Science Foundation, under Grant no. 782/08.

## 6    Conflict of Interest Statement

We wish to confirm that there are no known conflicts of interest associated with this publication and there has been no significant financial support for this work that could have influenced its outcome. No ethical approval is required.

**References**

1.    Tarbell, J.M., et al., *Fluid Mechanics, Arterial Disease, and Gene Expression.* Annual Review Of Fluid Mechanics, 2014. **46**: p. 591-614.

2. Davies, P.F., *Hemodynamic shear stress and the endothelium in cardiovascular pathophysiology.* Nat Clin Pract Cardiovasc Med, 2009. **6**(1): p. 16-26.
3. Reneman, R.S., T. Arts, and A.P. Hoeks, *Wall shear stress–an important determinant of endothelial cell function and structure–in the arterial system in vivo.* Journal of vascular research, 2006. **43**(3): p. 251-269.
4. Ku, D.N., et al., *Pulsatile flow and atherosclerosis in the human carotid bifurcation. Positive correlation between plaque location and low oscillating shear stress.* Arteriosclerosis, Thrombosis, and Vascular Biology, 1985. **5**(3): p. 293-302.
5. Shimogonya, Y., et al., *Can temporal fluctuation in spatial wall shear stress gradient initiate a cerebral aneurysm? A proposed novel hemodynamic index, the gradient oscillatory number (GON).* Journal of Biomechanics, 2009. **42**(4): p. 550-554.
6. Peiffer, V., S.J. Sherwin, and P.D. Weinberg, *Does low and oscillatory wall shear stress correlate spatially with early atherosclerosis? A systematic review.* Cardiovascular Research, 2013.
7. Kaspera, W., et al., *Morphological, hemodynamic, and clinical independent risk factors for anterior communicating artery aneurysms.* Stroke, 2014. **45**(10): p. 2906-2911.
8. Lu, X. and G.S. Kassab, *Nitric oxide is significantly reduced in ex vivo porcine arteries during reverse flow because of increased superoxide production.* The Journal of physiology, 2004. **561**(2): p. 575-582.
9. Moore, J.E., et al., *Fluid wall shear stress measurements in a model of the human abdominal aorta: oscillatory behavior and relationship to atherosclerosis.* Atherosclerosis, 1994. **110**(2): p. 225-240.
10. Caro, C., J. Fitz-Gerald, and R. Schroter, *Atheroma and arterial wall shear observation, correlation and proposal of a shear dependent mass transfer mechanism for atherogenesis.* Proceedings of the Royal Society of London. Series B. Biological Sciences, 1971. **177**(1046): p. 109-133.
11. Malek, A.M., S.L. Alper, and S. Izumo, *Hemodynamic shear stress and its role in atherosclerosis.* Journal of the American Medical Association, 1999. **282**(21): p. 2035-2042.
12. Resnick, N., et al., *Hemodynamic forces as a stimulus for arteriogenesis.* Endothelium, 2003. **10**(4-5): p. 197-206.
13. Tang, D., et al., *Image-based modeling for better understanding and assessment of atherosclerotic plaque progression and vulnerability: Data, modeling, validation, uncertainty and predictions.* Journal of Biomechanics, 2014. **47**(4): p. 834-846.
14. Finol, E.A. and C.H. Amon, *Flow-induced wall shear stress in abdominal aortic aneurysms: Part ii-pulsatile flow hemodynamics.* Computer Methods in Biomechanics & Biomedical Engineering, 2002. **5**(4): p. 319-328.
15. Le, T.B., I. Borazjani, and F. Sotiropoulos, *Pulsatile flow effects on the hemodynamics of intracranial aneurysms.* Journal of biomechanical engineering, 2010. **132**(11): p. 111009.
16. Boussel, L., et al., *Aneurysm growth occurs at region of low wall shear stress: patient-specific correlation of hemodynamics and growth in a longitudinal study.* Stroke, 2008. **39**(11): p. 2997-3002.
17. Kawaguchi, T., et al., *Distinctive flow pattern of wall shear stress and oscillatory shear index: similarity and dissimilarity in ruptured and unruptured cerebral* *aneurysm blebs: clinical article.* Journal of neurosurgery, 2012. **117**(4): p. 774-780.
18. Patti, J., F. Viñuela, and A. Chien, *Distinct trends of pulsatility found at the necks of ruptured and unruptured aneurysms.* Journal of neurointerventional surgery, 2014. **6**(2): p. 103-107.
19. Oyre, S., et al., *In vivo wall shear stress measured by magnetic resonance velocity mapping in the normal human abdominal aorta.* European Journal Of Vascular And Endovascular Surgery, 1997. **13**(3): p. 263-271.
20. Gharib, M. and M. Beizaie, *Correlation between negative near-wall shear stress in human aorta and various stages of congestive heart failure.* Annals of Biomedical Engineering, 2003. **31**(6): p. 678-85.
21. Poelma, C., et al., *Ultrasound imaging velocimetry: Toward reliable wall shear stress measurements.* European Journal of Mechanics-B/Fluids, 2012. **35**: p. 70-75.
22. Leow, C.H., et al., *Flow Velocity Mapping Using Contrast Enhanced High-Frame-Rate Plane Wave Ultrasound and Image Tracking: Methods and Initial in Vitro and in Vivo Evaluation.* Ultrasound in medicine & biology, 2015. **41**(11): p. 2913-2925.
23. Blake, J.R., et al., *A method to estimate wall shear rate with a clinical ultrasound scanner.* Ultrasound in medicine & biology, 2008. **34**(5): p. 760-774.
24. Womersley, J.R., *Method for the calculation of velocity, rate of flow and viscous drag in arteries when the pressure gradient is known.* The Journal of physiology, 1955. **127**(3): p. 553-563.
25. Tsangaris, S. and N. Stergiopulos, *The inverse Womersley problem for pulsatile flow in straight rigid tubes.* Journal of Biomechanics, 1988. **21**(3): p. 263-266.
26. Cezeaux, J.L. and A. van Grondelle, *Accuracy of the inverse Womersley method for the calculation of hemodynamic variables.* Annals of Biomedical Engineering, 1997. **25**(3): p. 536-546.
27. Niebauer, J. and J.P. Cooke, *Cardiovascular effects of exercise: role of endothelial shear stress.* Journal of the American College of Cardiology, 1996. **28**(7): p. 1652-1660.
28. Palatini, P. and S. Julius, *Elevated heart rate: a major risk factor for cardiovascular disease.* Clinical and experimental hypertension, 2004. **26**(7-8): p. 637-644.
29. Fox, K., et al., *Heart rate as a prognostic risk factor in patients with coronary artery disease and left-ventricular systolic dysfunction (BEAUTIFUL): a subgroup analysis of a randomised controlled trial.* The Lancet, 2008. **372**(9641): p. 817-821.
30. Böhm, M., et al., *Heart rate as a risk factor in chronic heart failure (SHIFT): the association between heart rate and outcomes in a randomised placebo-controlled trial.* The Lancet, 2010. **376**(9744): p. 886-894.
31. Yomosa, S., *Solitary waves in large blood vessels.* J. Phys. Soc. Japan, 1987. **56**: p. 506-520.
32. Nichols, W.W. and M.F. O'Rourke, *McDonald's Blood Flow in Arteries*, ed. t. ed.1998, London: Arnold.
33. Ku, D., *Blood flow in arteries.* Annual Review Of Fluid Mechanics, 1997. **29**: p. 399-434.
34. Gosling, R. and D.H. King, *Arterial assessment by Doppler-shift ultrasound.* Proceedings of the Royal Society of Medicine, 1974. **67**(6 Pt 1): p. 447.
35. Daidzic, N.E., *Application of Womersley Model to Reconstruct Pulsatile Flow From Doppler Ultrasound Measurements.* Journal of Fluids Engineering, 2014. **136**(4): p. 041102.

36. Evans, D., et al., *The relationship between ultrasonic pulsatility index and proximal arterial stenosis in a canine model.* Circulation Research, 1980. **46**(4): p. 470-475.
37. Bouillot, P., et al., *Multi-time-lag PIV analysis of steady and pulsatile flows in a sidewall aneurysm.* Experiments In Fluids, 2014. **55**(6): p. 1-11.
38. Martí–Fàbregas, J., et al., *Prognostic value of Pulsatility Index in acute intracerebral hemorrhage.* Neurology, 2003. **61**(8): p. 1051-1056.
39. Hashimoto, J. and S. Ito, *Pulse pressure amplification, arterial stiffness, and peripheral wave reflection determine pulsatile flow waveform of the femoral artery.* Hypertension, 2010. **56**(5): p. 926-933.
40. Formaggia, L., A. Quarteroni, and A. Veneziani, *Cardiovascular Mathematics: Modeling and simulation of the circulatory system*. Vol. 1. 2010: Springer Science & Business Media.
41. Thurston, G.B., *Viscoelastic properties of blood and blood analogs.* Advances in hemodynamics and hemorheology, 1996. **1**: p. 1-30.
42. Taylor, Z.J., et al., *Long-duration time-resolved PIV to study unsteady aerodynamics.* Instrumentation and Measurement, IEEE Transactions on, 2010. **59**(12): p. 3262-3269.
43. Zamir, M., *The Physics of Pulsatile Flow*2000: AIP Press/Springer-Verlag.
44. Leguy, C., et al., *Model-based assessment of dynamic arterial blood volume flow from ultrasound measurements.* Medical & Biological Engineering & Computing, 2009. **47**(6): p. 641-648.
45. van Grondelle, A. and J. Cezeaux, *Calculation of the Velocity Profile, Flow and Wall Shear Stress in Arteries From the Pressure Gradient: Importance of Distensibility and Taper*, in *Biofluid Mechanics*1990, Springer. p. 545-546.
46. Sahtout, W. and R.B. Salah, *Influence of the distensibility of large arteries on the longitudinal impedance: application for the development of non-invasive techniques to the diagnosis of arterial diseases.* Nonlinear biomedical physics, 2012. **6**(1): p. 2.

**List of Figures:**

Figure 3: Schematic description of the experimental setup: (a) hydraulic setup, including the computer-controlled two-directional pumping system, elastic tube embedded in a refractive index matched solution, and the pressure/discharge metering system. (b) Schematic view of the phase-averaging PIV measurement system, computer controlled waveform pressure signal used to trigger the PIV acquisition at each cycle.

Figure 4: PIV measured velocity profiles (symbols) and the corresponding inverse Womersley solution profiles (dashed-lines) for 14 runs at different time phases (color online).

Figure 5: Flow rate (dashed black lines, left axis) and the corresponding WSS values (red solid lines, right axis) as simulated for the 14 cases

Figure 6: An example of flow waveforms and respective WSS estimate for two representative experimental runs: run #10 with only positive flow throughout the entire cycle (left), and run #12, with total flow reversal (right) (a) Flow rate as a function of normalized time (Q(t/T)). Values integrated from the PIV measurements are shown by symbols and the associated error bars, while curves represent the values evaluated using the inverse Womersley model. (b) Comparison between WSS calculated from Womersley approximation (solid line) and WSS calculated from measured velocity profiles (dashed line with symbols).

Figure 7: Scatter plot of *OSI* versus Reynolds (a), Womersley (b) and PI (c) parameters. Regression between the measured OSI values and predicted *P%* values (d), based on the non-linear least-square fit of a second order multi-variable polynomial function f(Re,Wo,PI) . The cases in the graphs

Avrahami et al.　　　　　　　　　　　　　　　　　　　　　　　　　　　　　　　　　　26

are marked according to their run number (as listed in Table 1) (see legends at the bottom).

Figure 8: WSS parameters as a function of the pulsatility index (PI): (a) $P[\%]$ vs PI; (b) OSI vs PI. logarithmic trend lines were added to delineate correlation. Cases are marked by their run number.

**Table 1: Dimensionless parameters of the 14 runs**

| Run # | $Re_{max}$ | $\alpha$ | PI |
|---|---|---|---|
| 1 | 633 | 9 | 2.4 |
| 2 | 732 | 13 | 1.6 |
| 3 | 490 | 6 | 3.3 |
| 4 | 477 | 9 | 2.6 |
| 5 | 496 | 13 | 1.7 |
| 6 | 592 | 9 | 1.5 |
| 7 | 190 | 6 | 4 |
| 8 | 189 | 6 | 3.9 |
| 9 | 69 | 6 | 8.9 |
| 10 | 329 | 8 | 2.4 |
| 11 | 195 | 8 | 3.2 |
| 12 | 80 | 8 | 6.4 |
| 13 | 335 | 11 | 1.8 |
| 14 | 90 | 11 | 4.8 |

**Table 2: R-square values of the best fit curves of the WSS-based parameters versus the hemodynamic parameter**

| Parameter/Model | Re | $\alpha$ | PI | Re, $\alpha$ | Re, PI | Re, $\alpha$, PI |
|---|---|---|---|---|---|---|
| P% (Fig. 5) | 0.8 | 0.6 | 0.85 | 0.73 | 0.85 | **0.9** |
| min/max (Fig. 6) | 0.0 | 0.1 | 0.4 | 0.1 | 0.36 | **0.65** |
| OSI (Fig. 7) | 0.7 | 0.3 | 0.9 | 0.5 | 0.85 | **0.95** |